\documentclass[%
 reprint,
 amsmath,amssymb,
 aps,superscriptaddress,prb
]{revtex4-1}
\usepackage{braket}
\usepackage{graphicx}
\usepackage{dcolumn}
\usepackage{bm}
\usepackage{subfigure} 

\begin{document}
\title{A Quantum Model for Entropic Springs}
\author{Chiao-Hsuan Wang}
\affiliation{Joint Quantum Institute, College Park, MD 20742}
\affiliation{University of Maryland, College Park, MD 20742}
\affiliation{Joint Center for Quantum Information and Computer Science, College Park, MD 20742}
\author{Jacob M. Taylor}
\affiliation{Joint Quantum Institute, College Park, MD 20742}
\affiliation{Joint Center for Quantum Information and Computer Science, College Park, MD 20742}
\affiliation{National Institute
of Standards and Technology, Gaithersburg, MD 20899}


\begin{abstract}
Motivated by understanding the emergence of thermodynamic restoring forces and oscillations, we develop a quantum-mechanical model of a bath of spins coupled to the elasticity of a material. We show our model reproduces the behavior of a variety of entropic springs while enabling investigation of non-equilibrium resonator states in the quantum domain. We find our model emerges naturally in disordered elastic media such as glasses, and is an additional, expected effect in systems with anomalous specific heat and 1/f noise at low temperatures due to two-level systems that fluctuate.
\end{abstract}

\pacs{}
\maketitle
\section{Introduction}
The resonant, linear response of a variety of physical systems naturally leads to using harmonic oscillator approximations to describe their behavior. Recent breakthroughs in the fabrication and characterization of large mechanical oscillators have led to tremendous progress in exploring the quantum behavior of macroscopic systems \cite{OptoReview}, including reports of ground state cooling \cite{OConnell2010,Teufel2011,Chan2011,Coherent2012} and demonstrations of squeezed states beyond the standard quantum limit \cite{Squeeze2004}. However, not all that resonates is a quantum harmonic oscillator. A variety of materials exhibit elastic, Hooke's law-type behavior at high temperatures due to changes in their microscopic configurations, i.e., their entropy \cite{Meyer1932,Guth1934,Kuhn1942,Kratky1949,DNA1994,Titin1997}.
A typical example of such ``entropic springs" is given by a rubber band, where forces applied cause disordered strands to straighten, reducing their entropy and resulting in a restoring force near thermodynamic equilibrium.

Here we show how a resonator whose stiffness depends upon the configuration of rapidly thermalizing two-level systems (TLSs) is a natural example of an entropic spring. This type of system arises in a variety of atomic and condensed matter systems \cite{AHV,Phillips1972,TCdispersive,SpinBath,decohere}. Specifically, fast thermalization of the TLSs in such systems leads to the first quantum-mechanical model of an entropic spring. The corresponding quantum model is surprisingly simple and admits solutions in a variety of physically relevant scenarios. We consider a collection of TLSs coupled to the elasticity of a medium, and find this naturally satisfies the fast thermalization mechanism and results in an entropic spring restoring force.  Furthermore, entropic oscillations persist in the thermodynamic limit at any finite temperature, and we show numerically that in typical cryogenic environments mechanical resonators used in optomechanics have a non-vanishing entropic spring contribution. Nonetheless, quantum coherence persists in the thermodynamic limit.

\section{quadratic-coupled spin bath model}
As a simplified model, consider a 1D harmonic oscillator with a mass $M$ and a bare frequency $\omega_0$ interacting quadratically with a collection of independent TLSs. The Hamiltonian of the combined system is $H_S=H_0+V$ with
\begin{align}
H_0=\frac{p^2}{2M}+ \frac{M \omega_0^2x^2}{2} +\sum _{j=1}^N B \frac{\sigma _z^j+1}{2},
   \label{H0}
\end{align}
where the TLSs are described as pseudo spin-1/2 systems using Pauli matrices $\sigma_z^j$, $N$ is the number of independent TLSs, and $B$ is the energy difference of the TLSs.
Motivated by our analysis of the TLS defects below, we consider a coupling term between the oscillator and TLSs 
\begin{align}
V=\sum _{j=1}^N \frac{M \omega^2}{2 N} \frac{\sigma _z^j+1}{2}x^2.
   \label{V}
\end{align}
We denote this the quadratic-coupled spin bath model (QCSB), which is a also simplified form of strong coupling Tavis-Cummings model in dispersive region \cite{TCdispersive}. It can be pictured as a system whose effective spring constant is determined by the polarization $\left\langle \sigma \right\rangle=\sum_i\left\langle (\sigma_z^i+1)/2\right\rangle/N$. We are primarily interested in the thermodynamic limit $N \rightarrow \infty$. Here we consider homogeneous case for simplicity; distributions of parameters will be considered later.

The thermodynamic properties of the system are encoded in its partition function. Specifically, the quantum partition function of the QCSB model can be evaluated as $Z=\text{Tr} e^{-\beta H}$, $\beta=1/(k_B T)$ is the inverse temperature,
\begin{align}
Z_0 \approx Z_{\rm TLS}^N \frac{1}{2 \sinh \left(\frac{\beta \hbar}{2}\sqrt{\omega_0^2+\frac{\omega^2}{1+e^{\beta B}}}\right)} \equiv Z_{\rm TLS}^N Z_{\rm OSC}.
   \label{Z0}
\end{align}
We drop terms of $\mathcal{O}(\frac{1}{N})$ in the thermodynamic limit using the method of steepest descent. $Z_{\rm TLS}^N$ is the partition function of N independent TLSs, and $Z_{\rm OSC}$ can be interpreted as the quantum partition function of an oscillator with a temperature-dependent frequency $\omega_{\rm eff}=\sqrt{\omega_0^2+\omega^2 \left\langle \sigma \right\rangle_B}$, and polarization $\left\langle \sigma \right\rangle_B=1/(1+e^{\beta B})$.

We proceed to see whether the system responds to an external perturbation as an entropic spring. We add an external field $H_f=-fx$ corresponding to a constant force $f$. The partition function becomes 
$Z \approx Z_0 \exp(\beta f^2/2 M \omega_{\rm eff}^2) $ for small $f$.

The Helmholtz free energy under the perturbation is
$A=U-ST=-\ln(Z)/\beta=A_0-f^2/(2 M \omega_{\rm eff}^2)$, with unperturbed Helmholtz free energy $A_0=-\ln(Z_0)/\beta$.
Crucially, a portion of the change in A arises from the change in entropy of the TLSs, $S=-\frac{\partial A}{\partial T}=S_0-k_B\frac{\beta^2Be^{\beta B}\omega^2f^2}{2(1+e^{\beta B})^2M\omega_{\rm eff}^4}$; $S_0=-\frac{\partial A_0}{\partial T}$.

This entropy change arises microscopically by assuming that the spins thermalize much faster than the motion of the oscillator. Specifically, consider $x$ now as a constant. Under fast thermalization, each spin goes to a Gibbs state according to the instantaneous energy difference as \begin{align}
\rho_{\rm TLS}^j=\frac{1}{1+e^{-\beta (B +\delta x^2)}}\begin{pmatrix} e^{-\beta (B +\delta x^2)} & 0\\ 0 & 1 \end{pmatrix}.
   \label{Gibb}
\end{align}
Here $\delta=M\omega^2/2N.$ The total entropy of the spins is thus a function of $x$ expressed as
\begin{align}
S_{\rm TLS} \approx S_B-k_B\frac{\beta^2 M \omega^2 B e^{\beta B}}{2(1+e^{\beta B})^2}x^2,
   \label{Sofx}
\end{align}
where $S_B$ is the total entropy of spins with energy difference $B$. Adding a force $f$ to the system will displace the oscillator equilibrium position by an amount $\Delta x =f/(M\omega_{\rm eff}^2)$ and decrease its entropy by $\Delta S=-k_B\frac{\beta^2 M \omega^2 B e^{\beta B}}{2(1+e^{\beta B})^2}(\Delta x)^2=-k_B\frac{\beta^2Be^{\beta B}\omega^2f^2}{2(1+e^{\beta B})^2M\omega_{\rm eff}^4}$. The previously found entropy change is recovered.

From the first law of thermodynamics and eq.(\ref{Sofx}), there will be a corresponding static entropic force
\begin{align}
F_S=T \frac{\partial S_{\rm TLS}}{\partial x}=-\frac{M \omega^2  \beta B e^{\beta B}}{(1+e^{\beta B})^2}x \equiv M\omega_S^2 x.
   \label{FS}
\end{align}
Thus, a portion of Hooke's law arises entirely from reconfiguration of the TLSs. We note that this isothermal process is reversible. We define an entropic spring parameter $R_S$ to quantify the entropic contribution,
\begin{align}
R_S \equiv \frac{\omega_S}{\omega_{\rm eff}} =\frac{\omega \sqrt{\beta B e^{\beta B}/(1+e^{\beta B})^2}}{\omega_{\rm eff}},
\label{RS}
\end{align}
the ratio between the frequency associated with entropic Hooke's law (eq. \ref{FS}) and the effective frequency. 

\section{Entropic Spring Master Equation}

The proposed QCSB model behaves like a spring in static equilibrium. However, without thermalization of the spins, the model will not undergo entropic oscillations. Thus, a full quantum treatment necessarily involves the thermalization of the spins.

We now derive the entropic spring master equation under the conditions of fast spin thermalization. The below derivation follows a standard quantum optics approach \cite{Carmichael}. Taking the total Hamiltonian $H=H_S+H_R+H_{\rm SR}$, we have the Hamiltonian of the oscillator-spins system in the QCSB model:
\begin{align}
H_S=\frac{p^2}{2M}+ \frac{M \omega_0^2 x^2}{2} +\sum _{i=1}^N B \frac{\sigma _z^i+1}{2}+\sum _{j=1}^N \frac{M \omega^2}{2 N} \frac{\sigma _z^j+1}{2}x^2 \, ;
   \label{HS}
\end{align}
the Hamiltonian of the thermal reservoir consisting of independent bosonic modes $r_{\vec{k},\lambda,j}$ with frequencies $\omega_{\vec{k},\lambda,j}$ for each spin $j$:
\begin{align}
H_R=\sum_{\vec{k},\lambda,j}\omega_{\vec{k},\lambda,j}r^{\dagger}_{\vec{k},\lambda,j}r_{\vec{k},\lambda,j} \, ;
   \label{HR}
\end{align}
and the interaction between the spins and the reservoir:
 \begin{align}
H_{\rm SR}&=\sum_{j=1}^N\sum_{\vec{k},\lambda} \left(\kappa^{\ast}_{\vec{k},\lambda,j}r^{\dagger}_{\vec{k},\lambda,j}\sigma^{j}_{-}+\kappa_{\vec{k},\lambda,j}r_{\vec{k},\lambda,j}\sigma^{i}_{+}\right)\notag\\
& \equiv\sum_{j=1}^N\left(\Gamma_j^{\dagger}\sigma^{j}_{-}+\Gamma_j\sigma^{j}_{+}\right) \, .
   \label{HSR}
\end{align}
We remark that the spin-reservoir interaction is under the dipole approximation and the rotating wave approximation, appropriate for $B \gg \frac{M\omega^2}{2N}$, and the reservoir only interacts with the spins.

To focus on the system-reservoir interaction, we now change to the interaction picture. An operator in the interaction picture $\tilde{O}(t)$ is defined through the unitary transformation
 \begin{align}
\tilde{O}(t)=e^{i (H_S+H_R) t}O_S(t)e^{-i (H_S+H_R) t},
   \label{OI}
\end{align}
and the time derivative of the interaction operator becomes
\begin{align}
\dot{\tilde{O}}(t)=-i\left[\tilde{H}_{SR}(t),\tilde{O}(t)\right].
  \label{Odot}
\end{align}

Defining $\omega_{\Sigma}^2\equiv \omega_0^2+\sum_j \frac{\omega^2}{ N}\frac{\sigma_z^j+1}{2} $, the oscillator operators in the interaction picture can be solved as follows:
\begin{align}
\tilde{x}(t)&=x\cos(\omega_{\Sigma}t)+\frac{p}{M \omega_{\Sigma}}\sin(\omega_{\Sigma}t),\notag\\
\tilde{p}(t)&=p\cos(\omega_{\Sigma}t)-M \omega_{\Sigma} x \sin(\omega_{\Sigma}t).
\label{xandp}
\end{align}
Since we are working in the regime where spins will quickly thermalize to their equilibrium value at temperature $T$, $\sum_j \frac{\sigma_z^j+1}{2N} \approx \frac{1}{1+e^{\beta B}}+O(\frac{1}{N})$, $\beta=1/k_B T$, we can approximate $\tilde{x}(t)\approx \bar{x}(t) = x\cos(\omega_{\bar{\sigma}}t)+\frac{p}{M \omega_{\bar{\sigma}}}\sin(\omega_{\bar{\sigma}}t)$ using a spin-independent frequency $\omega_{\bar{\sigma}}^2=\omega_0^2+\omega^2/(1+e^{\beta B})$ when evaluating spin evolution (while the coupling strength between $x$ and each spin is already of the order of $\delta$). We now examine the spins and the bath in the interaction picture:

\begin{align}
\tilde{\sigma}^i_{-}(t)&=\sigma^i_{-}e^{-iBt}\mathcal{T}\{ e^{- i \delta \int_0^t  \bar{x}^2(t_1)dt_1} \},\notag\\
\tilde{\sigma}^i_{+}(t)&=\sigma^i_{+}e^{iBt}\mathcal{T}\{ e^{i \delta \int_0^t  \bar{x}^2(t_1)dt_1} \},\notag\\
\tilde{\Gamma}_j^{\dagger}(t)&=\sum_{\vec{k},\lambda,j} \kappa^{\ast}_{\vec{k},\lambda,j}r^{\dagger}_{\vec{k},\lambda,j}e^{i\omega_{k}t},\notag\\
\tilde{\Gamma}_j(t)&=\sum_{\vec{k},\lambda,j} \kappa_{\vec{k},\lambda,j}r_{\vec{k},\lambda,j}e^{-i\omega_{k}t}.
  \label{sigmagammai}
\end{align}

Let $\chi(t)$ be the density operator for $S \otimes R$, then $\rho(t)= \text{tr}_R \left[\chi(t)\right]$ is the  reduced density matrix describing the system only. Assume the interaction is turned on at t=0 and the total Hilberspace starts with an uncorrelated (product) state $\chi(0)=\tilde{\chi}(0)=\rho(0)\otimes R_0$. The intial state of the reservoir is taken to be a thermal equilibrium state at temperature $T$, with density matrix $R_0={\displaystyle\prod_{\vec{k},\lambda,j}} e^{-\beta \omega_{\vec{k},\lambda,j}r^{\dagger}_{\vec{k},\lambda,j}r_{\vec{k},\lambda,j}}(1-e^{-\beta \omega_{\vec{k},\lambda,j}})$.

Starting from the exact time evolution of the $S \otimes R$ density matrix in interaction picture
\begin{align}
\dot{\tilde{\chi}}(t)=-i\left[\tilde{H}_{SR},\chi(0)\right]-\int_0^t dt{'} \left[\tilde{H}_{SR}(t),\left[\tilde{H}_{SR}(t{'}),\tilde{\chi}(t{'})\right] \right].
  \label{chidot}
\end{align}
Assuming weak coupling between the reservoir and the spins, we first introduce $(i)$\textit{Born approximation}, so that $\tilde{\chi}(t)\approx\tilde{\rho}(t)\otimes R_0$.
The time evolution of $\tilde{\rho}(t)\equiv tr_R\left[\tilde{\chi}(t)\right]$ can then be evaluated through the Born approximation as
\begin{align}
\dot{\tilde{\rho}}(t)=-\int_0^t dt{'} tr_R\left\{\left[\tilde{H}_{SR}(t),\left[\tilde{H}_{SR}(t^{'}),\tilde{\rho}(t')R_0\right] \right]\right\}.
  \label{rhodot}
\end{align}

Plugged in the interaction Hamiltonian for our model, the time evolution equation eq.($\ref{rhodot}$) can be explictly written down as
\begin{widetext}
\begin{align}
\dot{\tilde{\rho}}(t)=-\sum_j\int_0^t dt{'} & \left\{\sigma_{-}^j\mathcal{T} \{e^{-i\delta   \int_0^t \bar{x}^2(t_1)dt_1}\}\sigma_{+}^j \mathcal{T} \{e^{i\delta \int_0^{t'} \bar{x}^2(t_1)dt_1}\} \tilde{\rho}(t)e^{-iB(t-t{'})}tr_R\left[R_0\tilde{\Gamma_j^{\dagger}}(t)\tilde{\Gamma_j}(t{'})\right]+h.c.\right. \notag\\
&-\sigma^j_+\mathcal{T} \{e^{i\delta \int_0^{t'} \bar{x}^2(t_1)dt_1}\}\tilde{\rho}(t')\sigma_-^j \mathcal{T}\{e^{-i\delta \int_0^t \bar{x}^2(t_1)dt_1}\}e^{-iB(t-t{'})}tr_R\left[R_0\tilde{\Gamma_j^{\dagger}}(t)\tilde{\Gamma_j}(t{'})\right]+h.c.\notag\\
&+\sigma_{+}^j \mathcal{T}\{e^{i\delta \int_0^t \bar{x}^2(t_1)dt_1}\} \sigma_{-}^j\mathcal{T}\{e^{-i\delta \int_0^{t'} \bar{x}^2(t_1)dt_1}\}\tilde{\rho}(t)e^{iB(t-t{'})}tr_R\left[R_0\tilde{\Gamma_j}(t)\tilde{\Gamma_j^{\dagger}}(t{'})\right]+h.c.\notag\\
&\left.-\sigma^j_-\mathcal{T}\{e^{-i\delta \int_0^{t'} \bar{x}^2(t_1)dt_1}\}\tilde{\rho}(t)\sigma_+^j\mathcal{T}\{e^{i\delta \int_0^t \bar{x}^2(t_1)dt_1}\}e^{iB(t-t{'})}tr_R\left[R_0\tilde{\Gamma_j}(t)\tilde{\Gamma_j^{\dagger}}(t{'})\right]+h.c. \right\}.
  \label{rhodote}
\end{align}
\end{widetext}

Assuming short time correlation between the reservoir and the spins, we now make the second approximation, $(ii)$ \textit{Markov approximation}, so that operators change slowly with time within the integral. We can then replace $\tilde{\rho}(t{'})$ by $\tilde{\rho}(t)$, $\mathcal{T} \{e^{-i\delta   \int_0^t \bar{x}^2(t_1)dt_1}\}$ by $e^{-i\delta \bar{x}(t) t}$, and $\mathcal{T} \{e^{-i\delta   \int_0^{t'} \bar{x}^2(t_1)dt_1}\}$ by $e^{-i\delta \bar{x}(t) t'}$ inside the formula, and extend the lower bound of the time integral from $0$ to $-\infty$.

After applying the above approximations, terms with $\sigma_+^j, \sigma_-^j$ on the same side of $\tilde{\rho}(t)$ resemble that of damped two-level atoms with the thermalization rate associated with a frequency $B+\delta \bar{x}^2(t)$ depending on $x$. Terms with $\sigma_+^j, \sigma_-^j$ on opposite sides of $\tilde{\rho}(t)$ can be expanded through a complete basis of $\bar{x}(t)$ for further calculation. Take the $\sigma_{+}^{j}\tilde{\rho}(t) \sigma_{-}^{j}$ term for example,

\begin{widetext}
\begin{align}
&\int_{-\infty}^t dt{'} \sigma^j_+e^{i\delta \bar{x}(t)t{'}}\tilde{\rho}(t)\sigma_-^je^{-i\delta x_t^2(t) t}e^{-iB(t-t{'})}tr_R\left[R_0\tilde{\Gamma_j^{\dagger}}(t)\tilde{\Gamma_j}(t{'})\right]\notag\\
=&\int_{-\infty}^t dt{'} \int dx_1 dx_3   \Ket{x_1(t)}\Bra{x_1(t)}\sigma_+^j\tilde{\rho}(t)\sigma_-^j\Ket{x_3(t)}\Bra{x_3(t)}e^{i\delta x_1^2(t) t{'}-i\delta x_3^2(t) t}e^{-iB(t-t{'})}tr_R\left[R_0\tilde{\Gamma_j^{\dagger}}(t)\tilde{\Gamma_j}(t{'})\right]\notag\\
=&\int_{0}^{\infty} d\tau \int dx_1 dx_3 \Ket{x_1(t)}\Bra{x_1(t)}\sigma_+^j\tilde{\rho}(t)\sigma_-^j\Ket{x_3(t)}\Bra{x_3(t)}e^{i\delta (x_1^2(t)-x_3^2(t))t} e^{-i(B+\delta x_1^2(t)) \tau}tr_R\left[R_0\tilde{\Gamma_j^{\dagger}}(t)\tilde{\Gamma_j}(t-\tau)\right].
  \label{+rho-}
\end{align}
\end{widetext}
In the last equality we perform a change of variable using $\tau=t-t{'}.$

Plugging in the expressions of reservoir correlation functions and evaluate the integral through standard master equation derivation approach, the time evolution equation becomes
\begin{align}
\dot{\tilde{\rho}}(t)&=-\sum_j \frac{1}{2} \left\{ \sigma^j_+ \sigma^j_- \gamma_{\bar{x}(t)} (\bar{n}_{\bar{x}(t)}+1) \tilde{\rho}(t) \right.
\notag\\ &+\tilde{\rho}(t)(\bar{n}_{\bar{x}(t)}+1)\gamma_{\bar{x}(t)} \sigma_+^j \sigma_-^j \notag\\
&-\sigma_-^j \gamma_{\bar{x}(t)}(\bar{n}_{\bar{x}(t)}+1) e^{-i\delta x^2(t) t} \tilde{\rho}(t) e^{i\delta x^2(t)t} \sigma_+^j \notag\\
&- \sigma_-^j e^{-i\delta \bar{x}(t) t} \tilde{\rho}(t)  e^{i\delta \bar{x}(t) t} \gamma_{\bar{x}(t)}(\bar{n}_{\bar{x}(t)}+1) \sigma_+^j\notag\\
&+ \sigma^j_- \sigma^j_+\gamma_{\bar{x}(t)} \bar{n}_{\bar{x}(t)} \tilde{\rho}(t)+\tilde{\rho}(t) \bar{n}_{\bar{x}(t)}\gamma_{\bar{x}(t)} \sigma_-^j \sigma_+^j\notag\\
&- \sigma_+^j \gamma_{\bar{x}(t)} \bar{n}_{\bar{x}(t)} e^{i\delta \bar{x}(t) t} \tilde{\rho}(t) e^{-i\delta \bar{x}(t)t}\sigma_-^j\notag\\
&-\left. \sigma_+^j e^{i\delta \bar{x}(t)t}   \tilde{\rho}(t)  e^{-i \delta \bar{x}(t) t} \gamma_{\bar{x}(t)}\bar{n}_{\bar{x}(t)}\sigma_-^j \right\}
\label{rhoIdotfin}
\end{align}

Here we have $\gamma_{\bar{x}(t)} \equiv 2\pi J(B+\delta \bar{x}(t))$, the spin thermalization rate associated with energy $B+\delta\bar{x}(t)$, and $\bar{n}_{\bar{x}(t)} \equiv \frac{e^{-\beta (B+ \delta \bar{x}(t))}}{1- e^{-\beta (B+\delta \bar{x}(t))}}$, the thermal occupation number associated with energy $B+\delta\bar{x(t)}$. $\bar{x}(t)$ is the interaction position operator and
$J(\omega)$ is the spectral density of the reservoir. We have taken Lamb-shift terms to be zero by appropriate redefinition of B.

Now we transform the time evolution equation back to the Schr\"odinger picture using
\begin{align}
\dot{\rho}(t)=-i\left[ H_S, \rho(t) \right]+ e^{-i H_S t}\dot{\tilde{\rho}}(t)e^{i H_S t}.
\label{itos}
\end{align}

The complicated time-dependence in eq.(\ref{rhoIdotfin}) disappears neatly when transformed back to the Schr\"odinger representation. When the thermalization rate is faster than the mechanical oscillation, we find that the dynamical evolution of the entropic spring model can be described by an \textit{entropic spring master equation} 
\begin{align}
&\dot{\rho}(t)=-i\left[ H_S, \rho(t) \right] 
-\sum_j \frac{1}{2} \left\{ \sigma^j_+ \sigma^j_-\gamma_{x^2} (\bar{n}_{x^2}+1) \rho(t)\right.\notag\\
&+ \rho(t)  (\bar{n}_{x^2}+1) \gamma_{x^2}\sigma^j_+ \sigma^j_- - \sigma_-^j  \gamma_{x^2}(\bar{n}_{x^2}+1) \rho(t)  \sigma_+^j \notag \\
&-\sigma_-^j   \rho(t)(\bar{n}_{x^2}+1)\gamma_{x^2} \sigma_+^j + \sigma^j_-  \sigma^j_+\gamma_{x^2} \bar{n}_{x^2} \rho(t)\notag\\
&+\rho(t)  \bar{n}_{x^2}  \gamma_{x^2} \sigma^j_-  \sigma^j_+
\left.-\sigma_+^j \gamma_{x^2} \bar{n}_{x^2}  \rho(t) \sigma_-^j -\sigma_+^j \rho(t) \bar{n}_{x^2}\gamma_{x^2} \sigma_-^j \right\}.
\label{esmaster}
\end{align}

Here we have the thermalization rate $\gamma_{x^2} \equiv 2\pi J(B+\delta x^2)$, and the thermal occupation number $\bar{n}_{x^2} \equiv \frac{e^{-\beta (B+ \delta x^2)}}{1- e^{-\beta (B+\delta x^2)}}$.

Note that the above master equation decouples to a Lindblad master equation for damped TLSs and the unitary evolution for a harmonic oscillator when taking the coupling $\omega$ to zero. In the massive limit, in which the oscillator position can be treated as a static constant, the spins will thermalize according to the probability
$\frac{P(\uparrow)}{P(\downarrow)}=e^{-\beta (B+\delta x^2)}$, which is consistent with the detailed balance as suggested by a static equilibrium.

Observables associated solely with the oscillator, i.e., operators which are functions of $x$ and $p$, will evolve according to the master equation (\ref{esmaster}) as
\begin{align}
\left\langle\dot{O}(x,p,t)\right\rangle&=\text{Tr}\left\{O\dot{\rho}(t)\right\}=-i\left\langle \left[O,  H_S\right] \right\rangle,
\label{xpevolution}
\end{align}
using the cyclic property of trace and the fact that $x$, and $p$ commute with spins. The overall contribution from the dissipation part vanishes. That is, $\left\langle x \right\rangle$ and $\left\langle p \right\rangle$ evolve only through the Hamiltonian. In the thermodynamic limit, we recover our static result: the oscillator evolves as a simple harmonic oscillator with an angular frequency $ \omega_{\rm eff}=\sqrt{\omega_0^2+\omega^2 \left\langle \sigma \right\rangle_B}$.

\section{Mesoscopic Effects}
\begin{figure}
\includegraphics[scale=0.38]{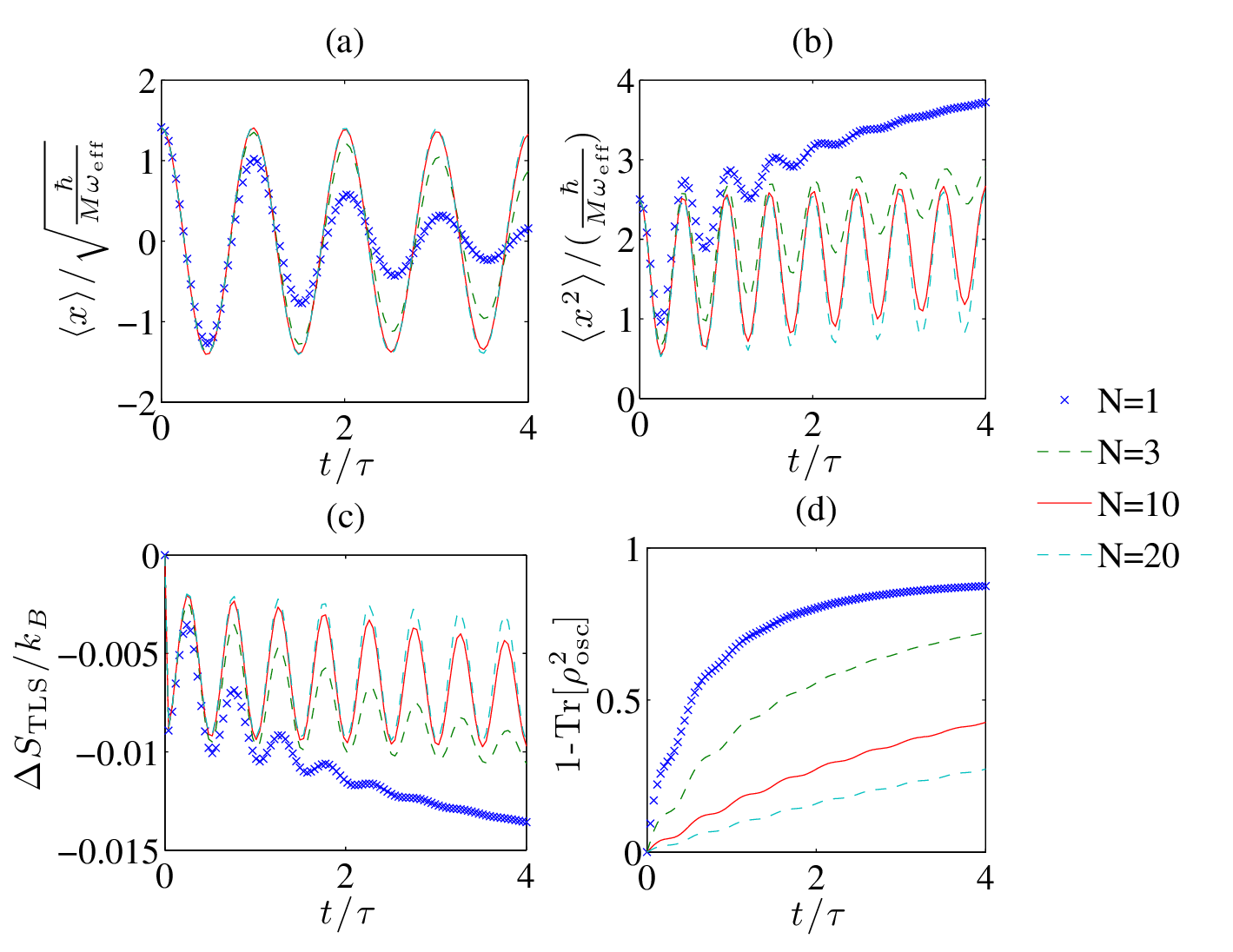}
\caption{\label{numerical}Numerical simulation based on the entropic spring master equation with parameters $\omega_0$= 0, $\omega= \sqrt{1+e^{\beta B}}$, $\gamma$= 10,  $\beta$= 0.01, $B$ = 100, and several $N$ as identified in the legend. $\tau$ is the oscillation period $\tau=2 \pi/\omega_{\rm eff}$. ($\hbar=M=\omega_{\rm eff}=1$)}
\end{figure}

\begin{figure}[h!]
\centering
\includegraphics[scale=0.3]{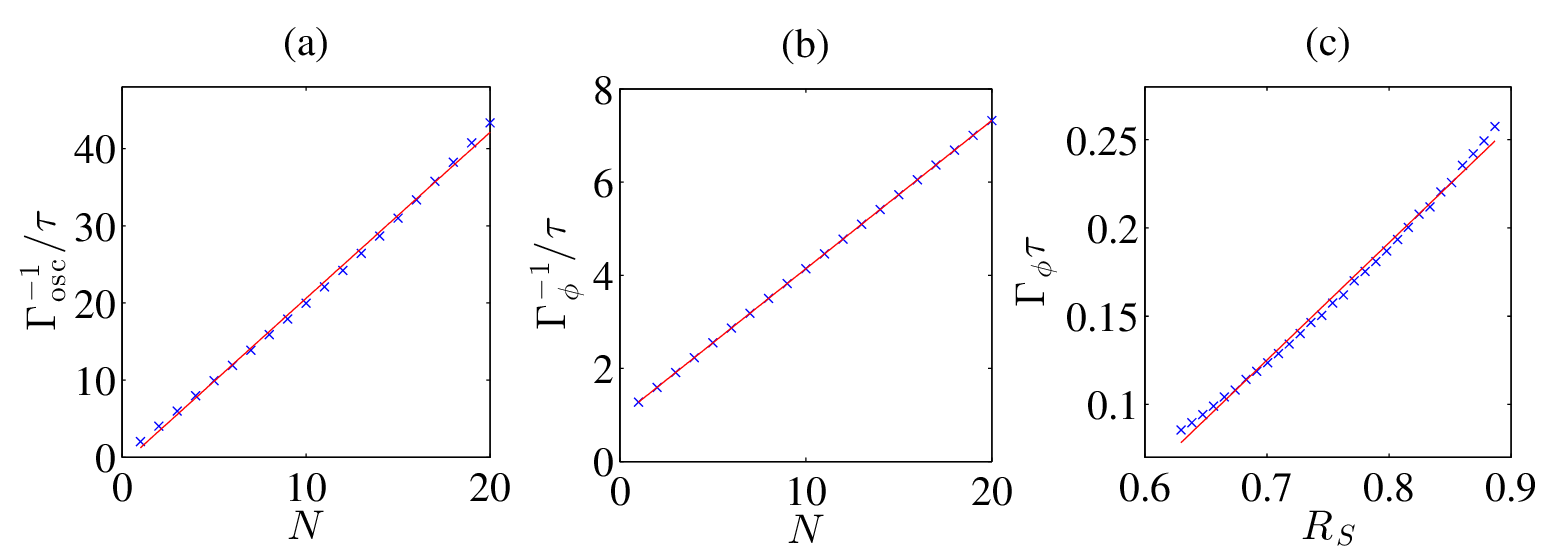}

\caption{(a) Inverse of position oscillation damping rate versus $N$. (b) Inverse of dephasing rate versus N. (c) Dephasing rate versus different entropic spring contribution.  We use the same set of parameters as Fig. \ref{numerical} for (a)(b), and use parameters $N = 10$, $\omega=\{0.7,0.71,0.72,\cdots,1\}$, $\omega_0=\sqrt{1-\omega^2}$, $\gamma$= 10,  $\beta$= 0.01, $B = 100$  for (c). $\tau$ is the oscillation period $\tau=2 \pi/\omega_{\rm eff}$. ($\hbar=M=\omega_{\rm eff}=1$) \label{qualitative}}
\end{figure}

We now consider mesoscopic corrections. The entropic spring master equation (eq.\ref{esmaster}) cannot be solved analytically due to nontrivial correlation between $x$ and $\sigma_z^j$ so we turn to numerical simulations. To overcome exponentially large density matrix representations, we use symmetric properties of the system Hamiltonian to reduce computational degrees of freedom. We define a set of elements $\left\lbrace C_j\right\rbrace $, j=0, 1, ..., N, symmetric in spins
\begin{align}
&C_0=I, C_1=\sum_{i=1}^N \sigma_z^i, C_2=\sum_{i=1}^N \sum_{j>i}\sigma_z^i \sigma_z^j \ , \cdots, \notag \\
&C_N=\sigma_z^1\sigma_z^2 \cdots\sigma_z^N .
   \label{Ci}
\end{align}
All completely symmetric density matrices of N spins can be expanded as $\rho_s=\sum_{j=0}^N a_jC_j$. We can then construct the transformation laws for $\sum_j \sigma_z^j C_k$, $\sum_j \sigma_+^j C_k \sigma_-^j$ and $\sum_j \sigma_-^j C_k \sigma_+^j$ and reduce the spin dimension from $2^N$ to N+1 with the above algebra.

Our numerical simulations (Fig.\ref{numerical}) follow the quantum trajectory of an initial state chosen as the Gibbs state of N spins with energy difference $B$ and a coherent state $\ket{\alpha}$ with $\alpha=1$ for the oscillator. For our simulation, we use units such that $\hbar=M=\omega_{\rm eff}=1$. We take $J(\omega) \sim$ constant near $B$ for simplicity, and set $\gamma$ large compared with $\omega_{\rm eff}$ to achieve fast thermalization and focus on cases with $\omega_0=0$ corresponding to oscillations arising entirely from reconfiguration of the coupled TLSs; cases with $\omega_0 \neq 0$ behave similarly.

Figure \ref{numerical} shows a simulation of the entropic spring evolution for several N. In Fig. \ref{numerical}(a)(b) we see damped oscillations in $\left\langle x(t)\right\rangle$ and $\left\langle x^2(t)\right\rangle$ with  damping rate suppressed as $N$ increases. Note that in contrast with eq. (\ref{xpevolution}), we see damping in oscillations due to uncertainties in the oscillation frequency caused by finite $N$. The entropic nature of the system is described by the TLSs von Neumann entropy $S_{\rm TLS}\equiv-\text{Tr}[\rho_{\rm TLS}\ln(\rho_{\rm TLS})]$, where $\rho_{\rm TLS}=\text{Tr}_{\rm osc}\rho$ is the reduced density matrix describing the TLSs. Fig .\ref{numerical}(c) shows the evolution of the TLSs entropy minus its initial thermal equilibrium value, $\Delta S_{\rm TLS}=S_{\rm TLS}-S_{\rm thermal}$, which oscillates with $-\left\langle x^2(t)\right\rangle$ through an oscillating amplitude independent of $N$. These results agree with our requirement that the spins are constantly rethermalizing according to the value of $x^2$, and $S_{\rm TLS}$ is changing with $x^2$ with a $N$-independent coefficient as suggested by eq.(\ref{Sofx}). Finally we monitor the dephasing by plotting the impurity of the oscillator, 1-Tr[$\rho_{\rm osc}^2(t)$], in Fig. \ref{numerical}(d), where $\rho_{\rm osc}(t)=\text{Tr}_{\rm TLS}[\rho(t)]$ is the oscillator reduced density matrix. The dephasing reduces as N increases.

The qualitative behavior of the damping of $\left\langle x \right\rangle$ and dephasing of 1-Tr[$\rho_{\rm osc}^2(t)$]  is followed in Fig. \ref{qualitative}. By fitting the evolution with an exponential decay envelope, the damping rate of $\left\langle x(t) \right\rangle$ is found to be proportional to 1/N (Fig. \ref{qualitative}(a)). The dephasing is also suppressed by $1/N$ as shown in Fig. \ref{qualitative}(b) and is proportional to the entropic spring parameter $R_S$ as shown in Fig. \ref{qualitative}(c).

The numerical simulation shows the intriguing result that,  in thermodynamic limit, the dephasing rate of the oscillator vanishes while amplitude of the oscillation of the entropy $S_{\rm TLS}$ stays constant. This phenomenon can be explained qualitatively by interpreting the thermal spins as a continuous weak measurement of $x^2$ through the coupling. The flipping rate of TLSs is $\Gamma_{\rm flip}=\gamma_{x^2}(\bar{n}_{x^2}+1)P(\uparrow)+\gamma_{x^2}\bar{n}_{x^2}(1-P(\uparrow)) \approx 2\gamma_{x^2} e^{-\beta (B+\delta x^2)}/(1-e^{-2\beta (B+\delta x^2)})$, assuming $\gamma_{x^2}$ is a constant and under fast thermalization approximation. The measurement rate is proportional to the flipping rate condition upon the change in $x^2$, which is of the order of $x^2\partial_{x^2}\Gamma_{\rm flip} \propto \sum_j \delta x^2 \sim \mathcal{O}(1)$, causes the N-independent entropy oscillation amplitude. The measurement back-action depends on $x^4$, therefore the dephasing rate is proportional to $\sum_j \delta^2 x^4 \sim \mathcal{O}(1/N)$, which vanishes in thermodynamic limit.

In principle, the high temperature limit of this model should be treatable in a semi-classical approximation where the quantum nature of the spin bath is largely ignored. However, such treatments would necessarily miss the mesoscopic dephasing effects due to the effective back action of the spin bath on the resonator shown in Fig. \ref{numerical} and Fig. \ref{qualitative}.

\section{Microscopic Origins: Two Level Systems in Amorphous Solids}

We now consider physical systems described by our model. Our QCSB entropic spring Hamiltonian follows naturally from the interaction between a single phonon mode and a collection of defects in amorphous solids. Defects in such materials can be microscopically modeled as atoms which can tunnel between two local ground states of asymmetric double-well potentials \cite{AHV,Phillips1972}. The effective Hamiltonian of a TLS defect is
\begin{align}
H_{\rm TLS}=\frac{\Delta}{2}\sigma_z-\frac{\Lambda}{2}\sigma_x=
\frac{1}{2}\begin{pmatrix} \Delta & -\Lambda\\ -\Lambda & -\Delta \end{pmatrix},
   \label{H0TLS}
\end{align}
where $\Delta$ is the asymmetry of the double-well, and $\Lambda$ is the tunneling between the wells. This model is used to explain anomalous low temperature specific heat behavior in glassy solids \cite{AHV}, and yields a 1/f noise spectrum in macroscopic properties such as electrical resistance \cite{1/f}.

An external strain field adds a perturbation to TLS Hamiltonian of the form \cite{Halperin1976}
\begin{align}
H_p=\frac{1}{2}\sum_{ij} \gamma_{ij} S_{ij}\begin{pmatrix} 1 & 0\\ 0 & -1 \end{pmatrix}  \equiv \frac{s}{2} \sigma_z,
   \label{H1TLS}
\end{align}
where $s=\sum_{ij} \gamma_{ij} S_{ij}$, $\gamma_{ij}$ are the coupling constants, and $S_{ij}$ is the strain field tensor. The displacement field caused by a single longitudinal wave can be expressed as quantized phonon creation and annihilation operators by $\vec{u}=\sqrt{\frac{\hbar}{2 \omega_k V \rho}}\left( \hat{a}^{\dagger}_k+\hat{a}_k\right)\cos(kz)\hat{z}$, where $\omega_k=v_Lk$, $k$ is the wave vector and $v_L$ is the sound velocity for longitudinal waves traveling in the solid. The strain field tensor is followed by $S_{ij}=(\partial_j u_i+\partial_i u_j)/2$. Assuming the interaction is isotropic, the coupling term is of the form $H_p= g \sigma_z \sqrt{\frac{\hbar}{2M\omega_p}}\left( \hat{a}^{\dagger}_k+\hat{a}_k\right)$ with coupling strength g. For transverse waves the above analysis holds  with $\gamma_L$ replaced by $\gamma_T$, and $v_L$ replaced by $v_T$.

\begin{figure}
\includegraphics[scale=0.4]{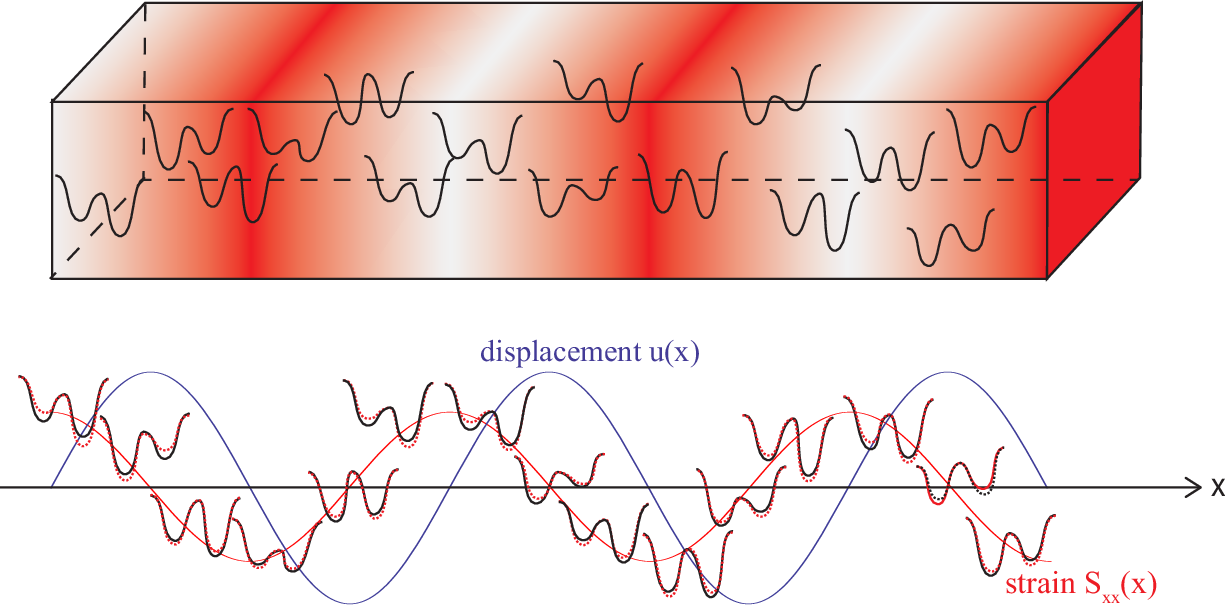}
\caption{\label{TLSphonon}A single longitudinal phonon mode interacts with a collection of two level system defects in a solid.}
\end{figure}

Changing to the diagonal basis of $H_{\rm TLS}$, the coupling between a single phonon mode with frequency $\omega_p$ and intrinsic TLS defects inside is of a general form
\begin{align}
H= \frac{p^2}{2M} +\frac{M\omega_p^2 x^2}{2}+\sum_j \frac{\epsilon_j }{2}\sigma_z^j+x\sum_j(g_{x,j} \sigma_x^j+g_{z,j}\sigma_z^j).
   \label{TLSH}
\end{align}
Here we have $x= \sqrt{\frac{\hbar}{2M\omega_p}}(a+a^{\dagger}),p=i\sqrt{\frac{\hbar M \omega_p}{2}}(a^{\dagger}-a) $, $\epsilon_j \equiv \sqrt{\Delta_j^2+\Lambda_j^2}$ is the excitation energy for the $j^{\rm th}$ TLS, $g_{x,j}=-g \Lambda_j/\epsilon_j$, and $ g_{z,j}=g\Delta_j/\epsilon_j$.

We are working in the physical region where the TLS energy differences $\epsilon$ is the dominant energy scale in the system, $\omega_p$, $g_x$, $g_z \ll \epsilon$. We first rotate the $j^{\rm th}$ spin about $\sigma_y^j$ with a small angle $\theta^j=xg_{x,j}/(xg_{z,j}+\epsilon_j/2)$ to diagonalize $\epsilon_j \sigma_z^j/2+x (g_{x,j}\sigma_{x}^j+g_{z,j}\sigma_{z}^j)$, which also boosts $p$ to $p-\sigma_y^j g_{x,j}/\epsilon_j+\mathcal{O}(\frac{1}{\epsilon_j^2})$. The generated $\sigma_y^j$ term in $p$ is negligible in a rotating-wave approximation since $\epsilon_j$ is the largest energy scale in the system.
   
We then perform a polaron-type transformation $U=e^{-i \xi p}$ to remove terms linear in $x$, in order to better approximate the dressed vacuum, with $\xi =\frac{\sum_j g_{z,j}\sigma_z^j}{M\omega_p^2+2\sum_j \frac{g_{x,j}^2}{\epsilon_j}\sigma_z^j}\approx \sum_j \frac{g_{z,j}}{M\omega_p^2}\sigma_z^j $
assuming the fluctuating TLSs have a minor effect on existing mechanical oscillation. 

Finally, we have
\begin{align}
&H'\approx \frac{p^2}{2M}+\frac{M(\omega_p^2-\bar{\omega}^2/2)}{2}x^2+\sum_j\frac{M \omega_{j}^2}{2}\frac{\sigma_z^j+1}{2}x^2 \notag\\ & + \sum_j \frac{\epsilon_j }{2} \sigma_z^j-\sum_{i,j} \frac{g_{z,i}g_{z,j}}{2 M \omega_p^2}\sigma_z^i\sigma_z^j
   \label{TLSfinal}
\end{align}   
Here $\bar{\omega}^2=\sum_j \omega_j^2=\sum_j \frac{4g_{x,j}^2}{M\epsilon_j}$, $M=V\rho$. Note that the spring constant contribution from each TLS is $\propto 1/M \propto 1/N$.
Our entropic spring Hamiltonian now arises naturally from this physical system, up to an extra Ising term. The strength of the Ising-like term is negligible compared to $\epsilon_j$, and is thus omitted in the thermodynamic analysis. 

\begin{figure}
\includegraphics[scale=0.5]{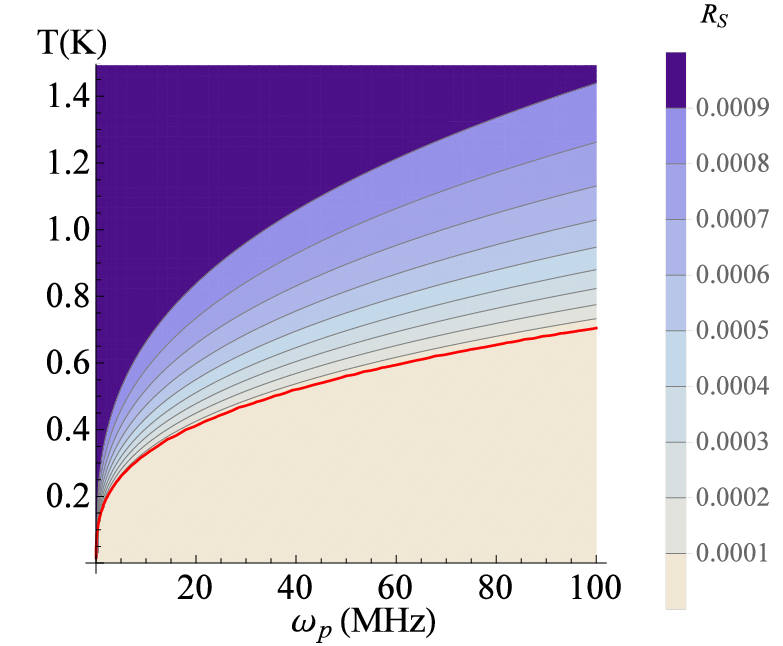}
\caption{\label{weff}Entropic spring parameter (Eq.~\ref{RSphysical}) as a function of resonator frequency $\omega_p$ and bulk temperature $T$ for parameters in the text corresponding to a BAW resonator in $\text{SiO}_2$ similar to Ref.~\cite{Tobar}. At intermediate temperatures around a tenth of a percent of oscillation is entropic in nature.}
\end{figure}

Based on the above, we can estimate the entropic spring parameter. Unlike the uniform case, as our initial QCSB model, we have to sum over the distribution of TLSs physical parameters. The entropic spring parameter $R_s$ is
\begin{align}
R_s=\sqrt{\frac{\sum_j\frac{4 g_{x,j}^2}{M \epsilon_j}\frac{\beta \epsilon_j e^{\beta \epsilon_j}}{(1+e^{\beta \epsilon_j})^2}}{\omega_p^2-\sum_j \frac{4 g_{x,j}^2}{M \epsilon_j} \frac{1}{2} \tanh{\frac{\beta \epsilon_j}{2}}}}.
   \label{RSphysical}
\end{align}

   For a standard tunneling TLS model that reproduces experimental 1/f noise, the probability distribution function for a TLS with energy difference $\epsilon$ and relaxation rate $\Gamma$ can be expressed as $P(\epsilon,\Gamma)=\bar{P}/(2\Gamma\sqrt{1-\Gamma/\Gamma_{\rm max}(\epsilon)})$ \cite{Jackle1972}. Here we have a constant density of states $\bar{P}$, and the TLS relaxation rate $\bar{P}$ as
$\Gamma(\epsilon,\Lambda)=\left(\frac{\gamma_L^2}{v_L^5}+2 \frac{\gamma_T^2}{v_T^5}\right)\frac{\epsilon \Lambda^2}{2 \pi \rho \hbar^4}\coth(\beta \epsilon /2)$ \cite{Jackle1972,Black1978}. The maximum relaxation rate for an energy difference $\epsilon$ is
$\Gamma_{\rm max}(\epsilon) \equiv \Gamma(\epsilon,\epsilon)$ since $\Lambda \leq \epsilon.$

The summation over TLSs is equivalent to integration over probability distribution, $\sum_j\rightarrow V \int d \epsilon \int d \Gamma P(\epsilon,\Gamma).$ At temperature T, the contributing TLSs are those of $\epsilon \lesssim k_B T$, since the entropic spring model works in the physical region that the TLSs fluctuate, which sets $\epsilon_{\rm max}$ to $k_B T$. To be consistent with previous analysis, we consider $\Gamma>\omega_p$, which sets the lower limit of $\Gamma$ to $\omega_p$ and $\epsilon_{\rm min}$ such that $\Gamma_{\rm max}(\epsilon_{\rm min})=\omega_p$.

By assuming small displacements, we approximate the sinusoidal displacement field to a triangle wave, and the variation of $\cos(kx)$ in x is now a constant slope $\frac{k}{\pi}$ up to a $\pm $ sign. For a phonon polarization denoted by $\alpha$, using the fact that $\Lambda^2/\epsilon^2=\Gamma/\Gamma_{max}(\epsilon)$, $g_{x,j}^2$ is identified as $g_{x,j}^2=\gamma_{\alpha}^2 \hbar\omega_p\Gamma /4\pi^2 v_{\alpha}^2  V \rho \Gamma_{max}(\epsilon)$. The entropic spring parameter can be evaluated through the integration over the above identified parameters.

There has been recent success in trapping longitudinal phonons in quartz Bulk Acoustic Wave (BAW) resonators with high quality factors \citep{Tobar}. For the case of $\text{SiO}_2$, we plot the lower bound of $R_S$ as a function of longitudinal mechanical frequency $\omega_p$ and temperature $T$ (Fig.\ref{weff}), using the parameters given in reference \cite{Black1978}. The maximum value of $R_S$ is about $0.1\%$, and there is no entropic contribution for $\Gamma_{\rm max}(k_BT)\lesssim \omega_p$, which happens when we have $2.86\times10^8 T^3K^{-3}s^{-1}<\omega_p$, and can be identified by the region below the red line in Fig. \ref{weff}.

\section{Outlook}
Our quantum entropic spring model predicts non-trivial contributions to the spring constant in disordered systems. However, direct observation of dephasing induced by the effect remains challenging. Future efforts in synthetic versions of this model, such as in the Tavis-Cummings interaction of an ensemble of atoms in an optical cavity, may allow experimental observation of the effects if an appropriate fast thermalization mechanism of the atoms can be provided.

\begin{acknowledgments}
We thank B. Halperin and Z.-X. Gong for helpful discussions. Funding is provided by NSF Physics Frontier Center at the JQI.
\end{acknowledgments}


\end{document}